\documentclass{PoS}
\usepackage{graphicx}
\usepackage{epsfig}
\usepackage{wrapfig}
\usepackage[small,bf]{caption}

\newcommand{\Wuppertal}{Bergische Universit\"at Wuppertal, Theoretical Physics, 42119 Wuppertal, Germany.}
\newcommand{\Budapest}{E\"otv\"os University, Theoretical Physics, P\'azm\'any P. S 1/A, H-1117, Budapest, Hungary.}
\newcommand{\Regensburg}{Institute for Theoretical Physics, Universit\"at Regensburg, D-93040 Regensburg, Germany.}
\newcommand{\Juelich}{J\"ulich Supercomputing Center, Forschungszentrum J\"ulich, D-52425 J\"ulich, Germany}
\newcommand{\Lambdamsbar}{\Lambda_{\overline{\rm MS}}}

\title{Lattice SU(3) thermodynamics and the onset of perturbative behaviour}
\ShortTitle{Lattice SU(3) thermodynamics and the onset of perturbative behaviour}
\date{\today}
\author{\speaker{Sz.~Bors\'anyi}$^1$,
G.~Endr\H odi$^{2,3}$,
Z.~Fodor$^{1,3,4}$,
S.~D.~Katz$^{3}$,
K.~K.~Szab\'o$^1$
}

\footnotetext[1]{\Wuppertal}
\footnotetext[2]{\Regensburg}
\footnotetext[3]{\Budapest}
\footnotetext[4]{\Juelich}

\abstract {
We present the equation of state (pressure, trace anomaly, energy density and
entropy density) of the SU(3) gauge theory from lattice field theory in an
unprecedented precision and temperature range. We control both finite size and
cut-off effects.  The studied temperature window ($0.7\dots 1000 T_c$)
stretches from the glueball dominated system into the perturbative regime,
which allows us to discuss the range of validity of these approaches. 
From the critical couplings on fine lattices we get $T_c/\Lambdamsbar=1.26(7)$
and use this ratio to express the perturbative free energy in $T_c$ units.
We also determine the preferred renormalization scale of the Hard Thermal Loop
scheme and we fit the unknown $g^6$ order perturbative coefficient at extreme
high temperatures $T>100T_c$.  We furthermore quantify the nonperturbative
contribution to the trace anomaly using two simple functional forms.
}

\FullConference{The XXVIII International Symposium on Lattice Field Theory \\
                June 14 - 19, 2010\\
                Viiasimius, Sardinia, Italy.}

\begin{document}

\section{Introduction}
The general feature of asymptotic freedom makes weak coupling approaches
very natural in non-abelian gauge theories, such as the SU(3) model, which
describes the gluonic degrees of freedom of Quantum Chromodynamics. At
asymptotically high temperatures low orders of perturbation theory may 
be acceptable, but at any lower scale that could be probed by a realistic
experiment an extension is necessary: either by the inclusion of very high order
diagrams, or by an efficient resummation scheme, such as Hard Thermal Loops (HTL).
Note however that analytic perturbative expansions are plagued by
infrared divergences due to which the series can be computed only up
to a given finite order.
There is strong simulation evidence that at low temperatures
($T_c\sim 260$ MeV)
the gluonic matter freezes and a first order transition takes
place. At even lower temperatures colorless non-perturbative excitations govern
the thermodynamics. To describe the phase transition or the glueball gas no
weak coupling scheme succeeds and one has to rely on a natively non-perturbative
approach, such as lattice field theory.

The past year witnessed considerable achievements on the side of the analytical
results. The HTL scheme has been used to calculate the pure
SU(3) gauge theory's thermodynamic potential to the next-to-next-to-leading
order (NNLO)~\cite{Andersen:2010ct}. The authors used their
results at intermediate temperatures ($\sim 4T_c$) where existing lattice data
were available. Later the same authors have extended their results to full QCD
(with massless quarks)~\cite{Andersen:2010wu} and 
found good agreement with the lattice data of the Wuppertal-Budapest collaboration
\cite{Borsanyi:2010cj} from about 300 MeV.


In conventional perturbation theory even higher orders can be computed by
dimensional reduction. The full expression up to  $g^6\log(g)$ order is given in~\cite{Kajantie:2002wa}
and was compared to the Bielefeld lattice data \cite{Boyd:1996bx} at $T=4.5T_c$. Fitting the
pressure (thermodynamic potential) the slope of the pressure curve was succesfully
predicted. This raised hope that at this high order perturbation theory
does possess some predictive power at phenomenological temperatures. In this work
we repeat this fitting procedure at a much higher temperature, where the sixth
order can be shown to be a minor correction to the fifth order.
Instead of treating the soft sector strictly perturbatively a screened
perturbation theroy can be formulated for the dimensionally reduced theory,
resulting in significantly better convergence of the free energy
\cite{Blaizot:2003iq}.

For more than a decade the renowned paper by Boyd et al~\cite{Boyd:1996bx} has
been the reference lattice simulation of the SU(3) theory in the temperature
range of $1\dots4.5 T_c$. It uses the plaquette gauge action at up to $N_t=8$
lattice spacing and an aspect ratio of 4. Here $N_t$ denotes the number of
lattice points in the Euclidean time direction, meaning that the lattice spacing
at any given temperature $T$ is $a=1/(TN_t)$. The fixed $N_t$ approach has been
introduced in Ref.~\cite{Engels:1990vr} and this work follows it, too. It
implies that the lattice spacing varies with temperature.
Continuum limit is achieved by performing an $1/N_t\to0$ extrapolation on the
data at a set of fixed physical temperatures.  The aspect ratio $r=LT$ sets the
ratio between space and time-like lattice points.

Since the publication of~\cite{Boyd:1996bx} several similar simulations were performed to study pure gauge theory. The equation of state has been recalculated using the Symanzik improved
gauge action~\cite{Beinlich:1997ia}.
This set of simulations have been further generalized
to SU($N_c$) theories with $N_c>3$ in Refs.~\cite{Panero:2009tv,Datta:2009jn}.
Alternatively, the equation of state can also be calculated by fixing the lattice
spacing, and using $N_t$ for tuning the temperature~\cite{Umeda:2008bd}. This
approach is mostly advantageous with Wilson-type dynamical fermions, and less
economic for the pure gluonic theory.

In most fixed $N_t$ simulation projects, like Ref.~\cite{Boyd:1996bx}, the
aspect ratio is kept constant to allow the use of a single lattice geometry.
This means that higher temperatures are simulated at smaller volumes.
Length scales that are present in (resummed) perturbative calculations, such as
$\sim T$, $\sim gT$ and $\sim g^2T$ are normally well accommodated in the
lattice, since the renormalized coupling $g$ drops only logarithmically as the
temperature increases.  Yet, to establish the range of validity of the
perturbative approach itself, one has to simulate the non-perturbative
$\sim T_c$ scale, too. In this sense, the aspect ratio sets the maximum
temperature as a precondition for the non-perturbativeness of the simulation:
$T\lesssim r T_c$.  In most previous works this was set to $r=4$.

\section{Simulation setup}

In this work we calculate the continuum equation of state of the SU(3) theory
using tree-level Symanzik improvement in the temperature range of
$T/T_c=0.7\dots 16$ (on $80^3\times 5$, $96^3\times 6$ and $114^3\times 7$
lattices). We also support this continuum extrapolation using an additional $N_t=8$ set of lattices ($64^3\times 8$) below $8 T_c$.
Furthermore we present a non-continuum data set that is valid up to
approximately $24T_c$ (on a $120^3\times 5$ lattice) and study finite volume effects using various smaller boxes. Finally, from our third set of
simulations we calculate the continuum equation of state
in a small box (on $40^3\times 5$, $48^3\times 6$ and $64^3\times 8$ lattices)
up to 1000 $T_c$.  These latter data we use to find the optimal free
parameters of existing perturbative calculations, e.g. the preferred
renormalization scale for the HTL scheme. The precision of our data points
exceeds any previous calculation by about an order of magnitude.

Although the lattice techniques for the analysis as well as for the generation
of lattice configurations have been well established, achieving the presented
statistics was a very challenging procedure. As explained in
Ref.~\cite{Engels:1990vr}, the normalized trace anomaly $(\epsilon-3p)/T^4$ 
is determined first. The normalized pressure $p/T^4$ and energy density
$\epsilon/T^4$ are calculated using thermodynamic relations. The trace anomaly
contains a quartic divergence, which is subtracted using lattice measurements of the same quantity (at the 
same lattice spacing) at a smaller temperature.
Because of this divergence the trace anomaly is rather difficult to measure.
Moreover, the value of $(\epsilon-3p)/T^4$ reduces rapidly with temperature as
one moves away from the transition region in either direction. 

Another challenging issue was the accurate determination of the non-perturbative
beta function corresponding to the Symanzik improved action.
Instead of using the string tension or the Sommer parameter, it was advantageous to define the lattice spacing in
terms of the transition temperature. To this end we determined the critical
couplings up to $N_t=20$ from the peak of the Polyakov loop susceptibility.
Matching to the universal two-loop running (in terms of the improved
coupling in the ``E'' scheme \cite{Bali:1992ru}, generalized for the case of
the Symanzik improved action) we determined the lambda
parameter in terms of the transition temperature:  $T_c/\Lambdamsbar=1.26(7)$.
(The error
is overwhelmingly systematic and reflects the sensitivity to various continuum
extrapolations.) This is consistent with the combination of previous determinations:
the Lambda parameter $\Lambdamsbar=0.614(2)(5) r_0^{-1}$  of~\cite{Gockeler:2005rv}
can be translated to $\sqrt{\sigma}$ units using $\sqrt{\sigma}r_0=1.192(10)$
(based on~\cite{Guagnelli:1998ud}) and then used with
$T_c/\sqrt{\sigma}=0.629(3)$ of Ref.~\cite{Boyd:1996bx}. Through our direct
result for $T_c/\Lambdamsbar$ one can easily translate the scale setting of
the perturbative expression for the SU(3) free energy to the lattice language.

Given the beta function one can simply relate the expectation value of
the gauge action $S_g$ (a weighted sum of $1\times1$ and $1\times2$
plaquettes) to $\epsilon-3p$~\cite{Engels:1990vr}.
In the standard lattice renormalization scheme this $\langle
S_g\rangle_T$ result is then subtracted from the value corresponding to zero
temperature. This would require a lattice with a large temporal size. In order
to be able to fit larger lattices into the given memory of our computer system
we used half-temperature subtraction, i.e. we calculated
$(\epsilon(T)-3p(T))/T^4-\frac{1}{16}(\epsilon(T/2)-3p(T/2)/(T/2)^4$
along the lines of our previous work \cite{Endrodi:2007tq}.
To finally
arrive at $(\epsilon(T)-3p(T))/T^4$ this partial result was supplemented with
another set of simulations at half temperature, double lattice spacing,
but same physical volume. For
these supplementary data we used the standard renormalization subtracting the
vacuum. The continuum limit from this combined technique is equal to what one
finds using the stardard scheme.

\section{Results}
\begin{figure}[ht!]
\centering
\vspace*{-0.2cm}
\hbox{
\epsfig{file=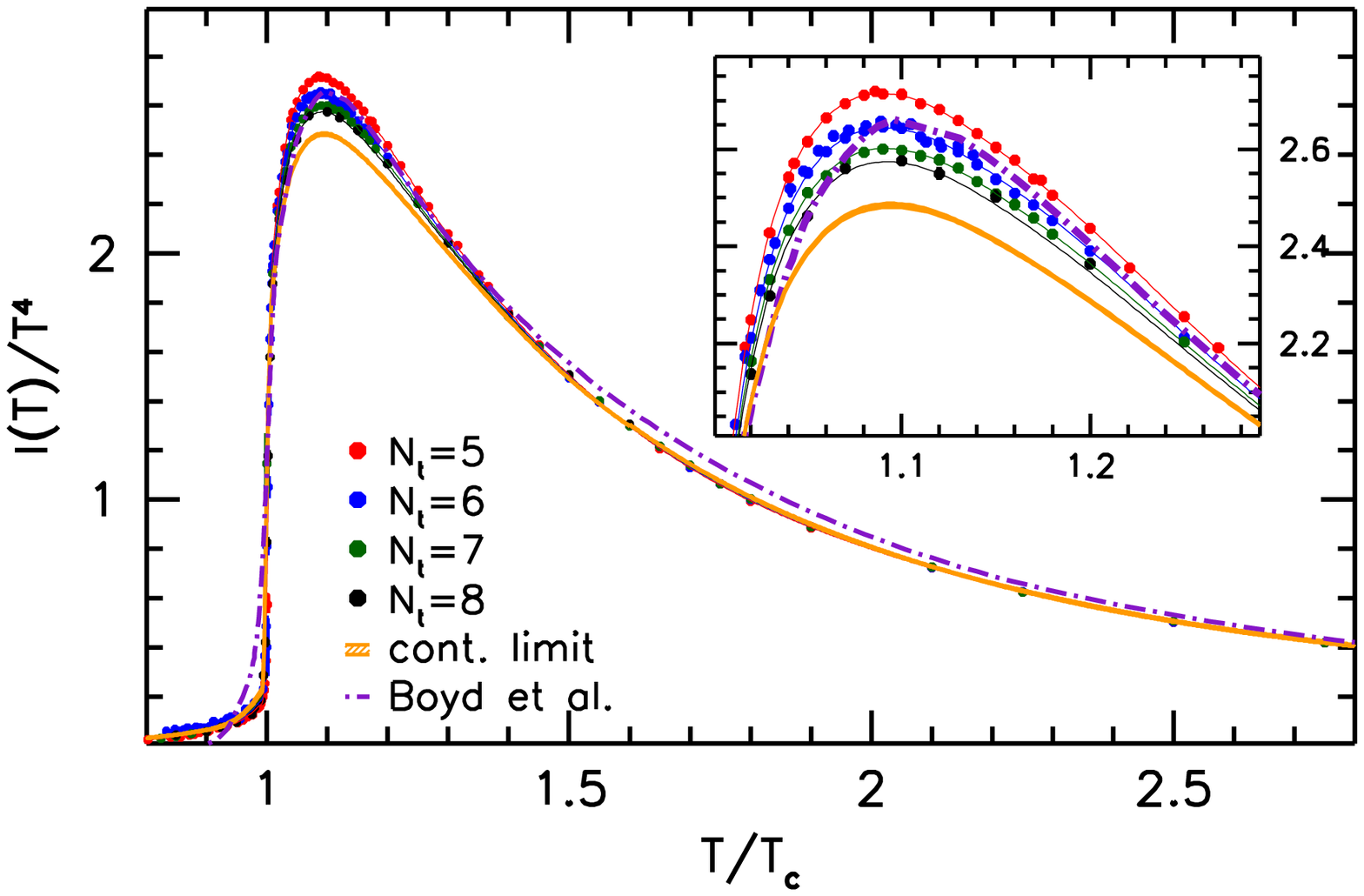,width=7.9cm,bb=18 360 592 718}
\epsfig{file=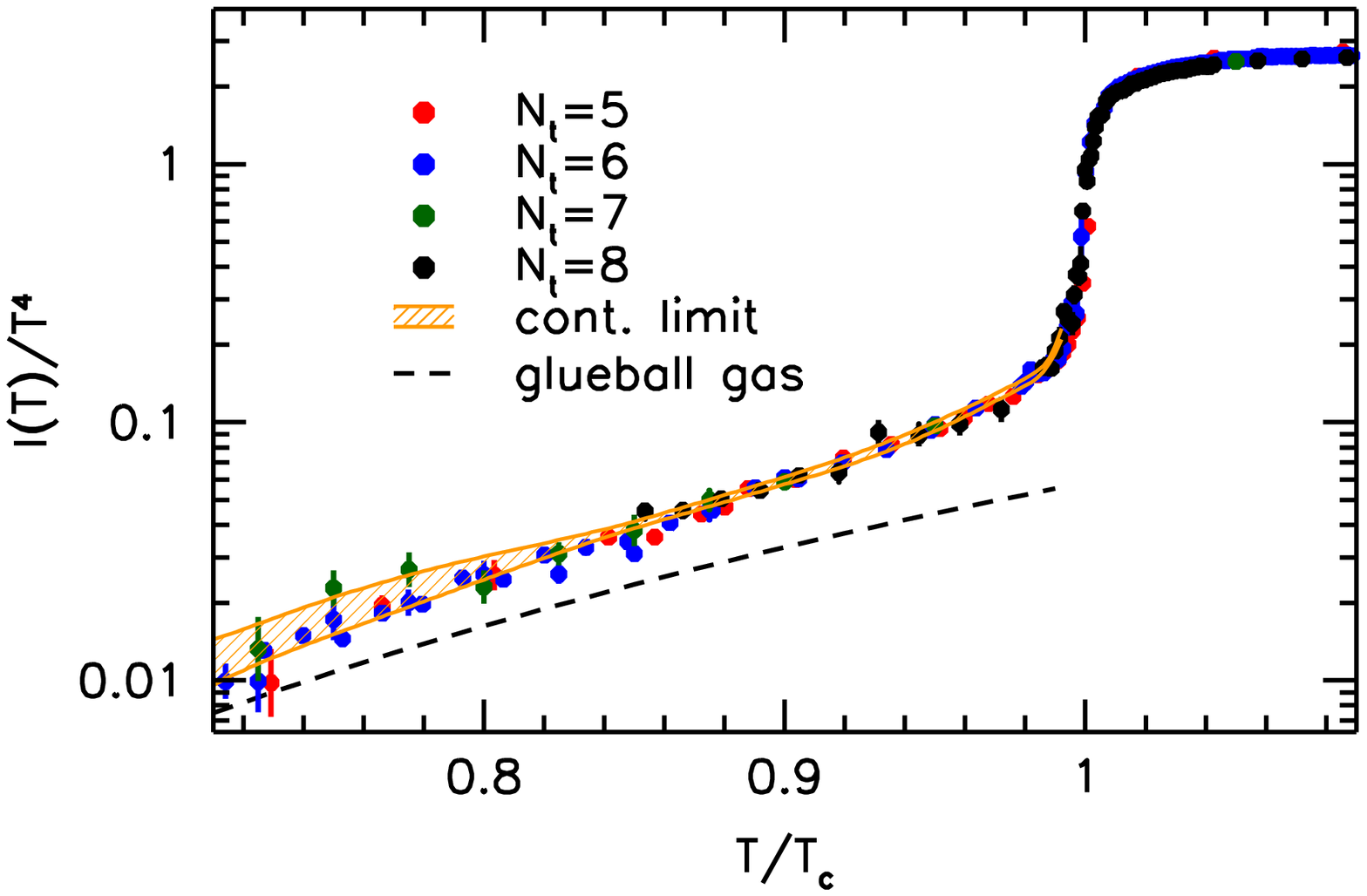,width=7.9cm,bb=18 360 592 718}
}
\vspace*{-0.1cm}
\caption{\label{fig:i_trans}
{\bf Left side. }The trace anomaly calculated on $80^3\times 5$, $96^3\times 6$, $114^3\times 7$ and $64^3\times 8$
lattices. The continuum extrapolation makes use of the $N_t^{-2}$ scaling of the Symanzik action. The data
deviates from Boyd et al~\cite{Boyd:1996bx} especially close and below the
transition.
{\bf Right side.} The trace anomaly in the confined phase measured on lattices with various lattice spacings and the continuum extrapolation (yellow band). The dashed line corresponds to
the gluon gas contribution, estimated from the twelve lightest glueballs.
}
\vspace*{-0.2cm}
\end{figure}

We start the presentation of the results with the reproduction of the Boyd et.
al. data~\cite{Boyd:1996bx} in the left side of figure~\ref{fig:i_trans}. We
fit our four large-volume data sets with different lattice spacings altogether,
using an $N_t$-dependent spline function.  As a result we have a smooth
function interpolating our data for each $N_t$ (colored lines in the figure),
together with a smooth, continuum extrapolated curve (yellow band in the
figure). The systematic error coming from this extrapolation procedure and the
statistical error are added in quadrature. 
We see a small discrepancy between our results and Ref.~\cite{Boyd:1996bx},
which might be attributed to the scale setting assuptions in \cite{Boyd:1996bx}.

Zooming in to the low temperure region we can explore the thermodynamics
of the confined phase. To find out to what extent glueballs dominate, we
calculated the trace anomaly contribution of the first twelve glueballs in
Ref.~\cite{Chen:2005mg} and plotted this together with our lattice results in
the right side of figure.~\ref{fig:i_trans}. It has been suggested in Ref.~\cite{Buisseret:2009eb} that
the apparent deficit between the lattice data might be explained if we allow
a temperature dependence for the glueballs. This dependence has already been determined for the 0++ and 2++ states~\cite{Ishii:2002ww}. This point was raised
when only a couple of inprecise simulation points existed below $T_c$ from
Panero's data set~\cite{Panero:2009tv}. Our data is in fairly good agreement
with this scenario, however, further studies are needed to understand the
question in more detail.

\begin{figure}[ht!]
\centering
\vspace*{-0.5cm}
\hbox{
\includegraphics*[height=5.45cm]{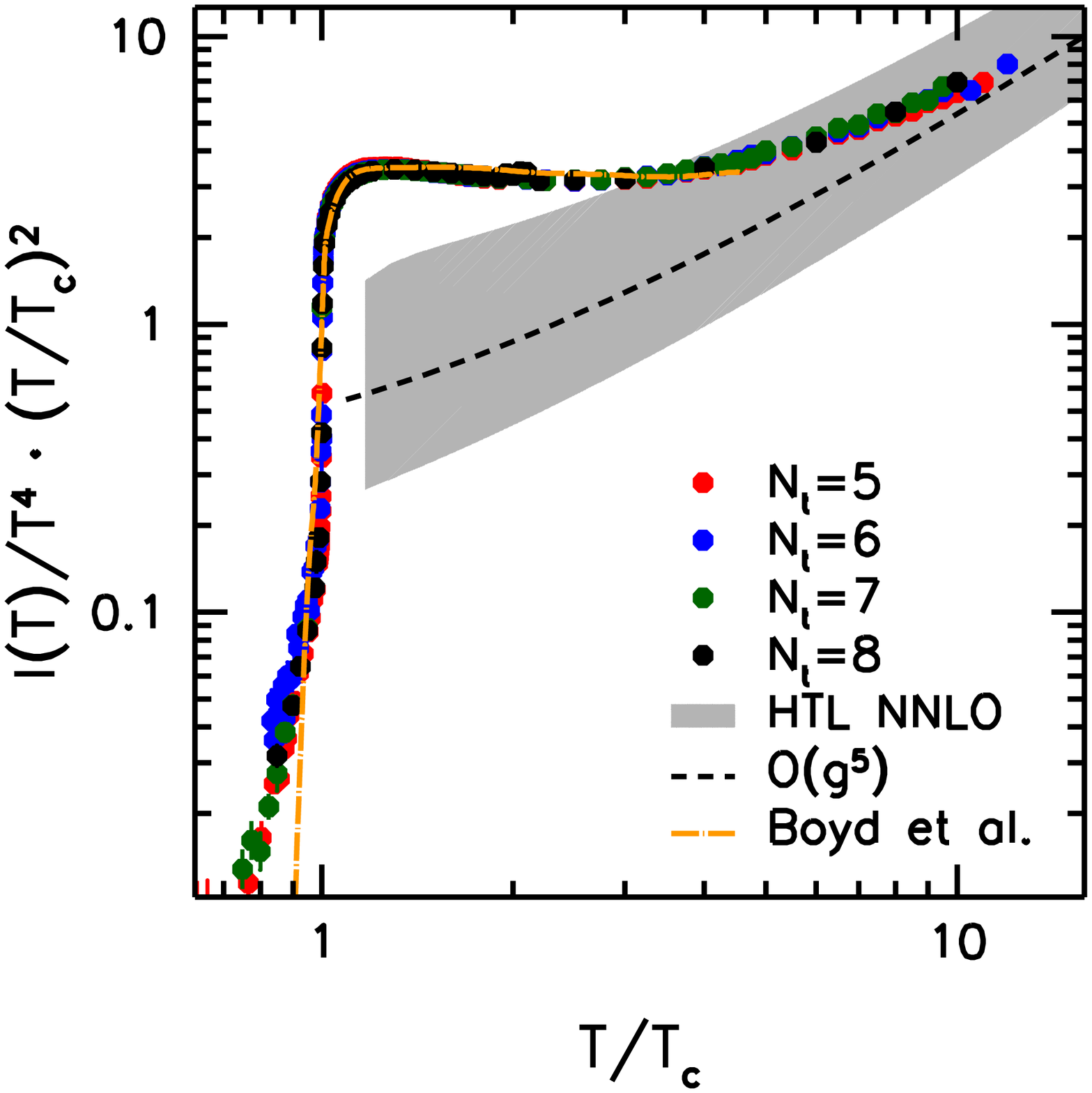}
\epsfig{file=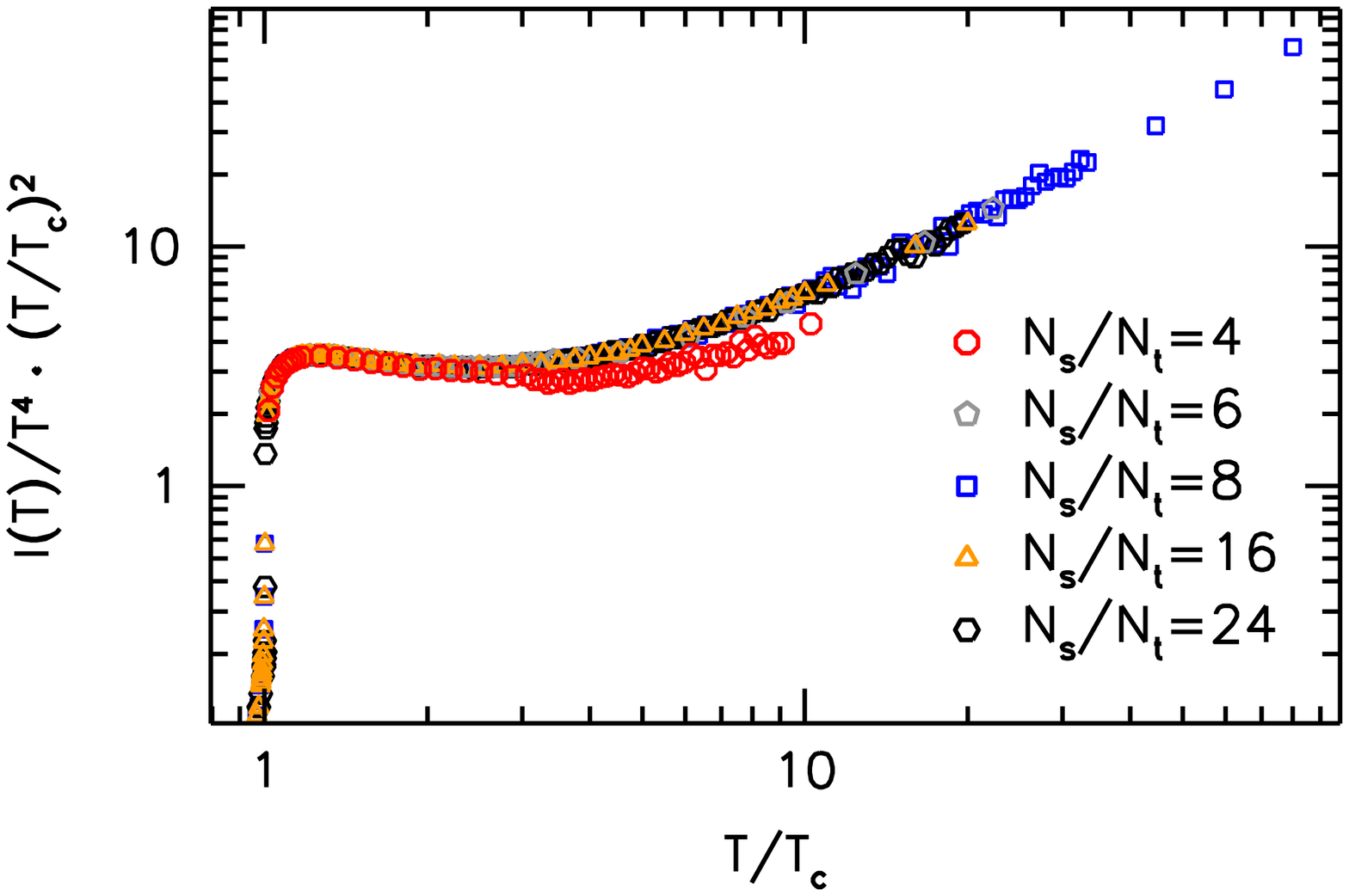,height=5.7cm,bb=18 350 592 718}
}
\vspace*{-0.2cm}
\caption{\label{fig:onset}
{\bf Left side.} Our results for the normalized trace anomaly multiplied by $T^2/T_c^2$ for $N_t=5,6,7$ and $8$ (red, green and blue dots, respectively). Also plotted are lattice results of~\cite{Boyd:1996bx}, $g^5$ perturbation theory~\cite{Kajantie:2002wa} and the HTL approach~\cite{Andersen:2010ct}.
{\bf Right side.} Trace anomaly with various volumes at one of our lattice resolutions $N_t=5$.
Unless the box is very small (e.g. $r=4$), there is no significant
difference whether or not the box size allows contributions from the inverse
$T_c$ scale.
}
\vspace*{-0.3cm}
\end{figure}

The strong non-perturbative effects in the equation of state can be emphasized
by plotting $(\epsilon-3p)/T^2$ instead of normalizing it to $T^4$
(left side of figure~\ref{fig:onset}). For dimensional reasons, any finite order perturbative
formula will only contribute logarithmic corrections to the $p(T)\sim T^4$
Stefan-Boltzmann law. Now plotting $(\epsilon-3p)/T^2$ we expect to see a curve
$\sim T^2$ times logarithmic corrections.  What lattice data, actually, does
show, is an approximately linear section up to $\sim 4 T_c$ temperature,
which then connects to the perturbative estimate. This linear segment was
more striking in the Bielefeld lattice data~\cite{Boyd:1996bx}, since the somewhat
bigger errors hid a more complex structure. Moreover the latter data set ended
at a temperature where the non-perturbative behaviour was still dominant.
The observed non-perturbative pattern induced speculations on a ``fuzzy'' bag model~\cite{Pisarski:2006yk} and potential explanations in terms of a dimension-2 gluon
condensate emerged~\cite{Pisarski:2000eq,Kondo:2001nq}.

In this work we do not go into the viability of various explanations to
the apparent $\sim T^2$ behaviour of $(\epsilon-3p)$, but we identify it 
as the dominant non-perturbative effect in the deconfined phase. Its effect
reduces at high temperatures and becomes unnoticeable from the lattice equation of
state, independently whether or not the lattice volume accommodates the inverse $T_c$
scale. One way of discussing the relevance of the $T_c$ scale
in the dynamics is to compare the trace anomaly at various volumes, as we do
in the right side of figure~\ref{fig:onset} for one lattice resolution. The ``standard'' aspect
ratio $r=4$ gives somewhat smaller values, but beyond $r\ge6$ our simulation
is not capable of resolving the difference.

We summarize our findings as i) the large volume lattice trace anomaly data shows qualitative (and as 
we find using the fitted $g^6$ order coefficient, also quantitative, see later) agreement with the perturbative
results from $T>10T_c$, and ii) we see no deviation between results
from various volumes (with $r\ge6$), moreover iii) the dominant non-perturbative
contribution loses significance as $\sim 1/T^2$.
These considerations suggest that -- even if the lattice volumes are ever shrinking as the temperature is increased -- our results are able to describe the physical trace anomaly 
(and its integral, the thermodynamic potential) within the error bars shown.
Of course, this assumes that all perturbatively
relevant scales are properly accounted for.  

\begin{figure}[ht!]
\centering
\vspace*{-0.6cm}
\hbox{
\epsfig{file=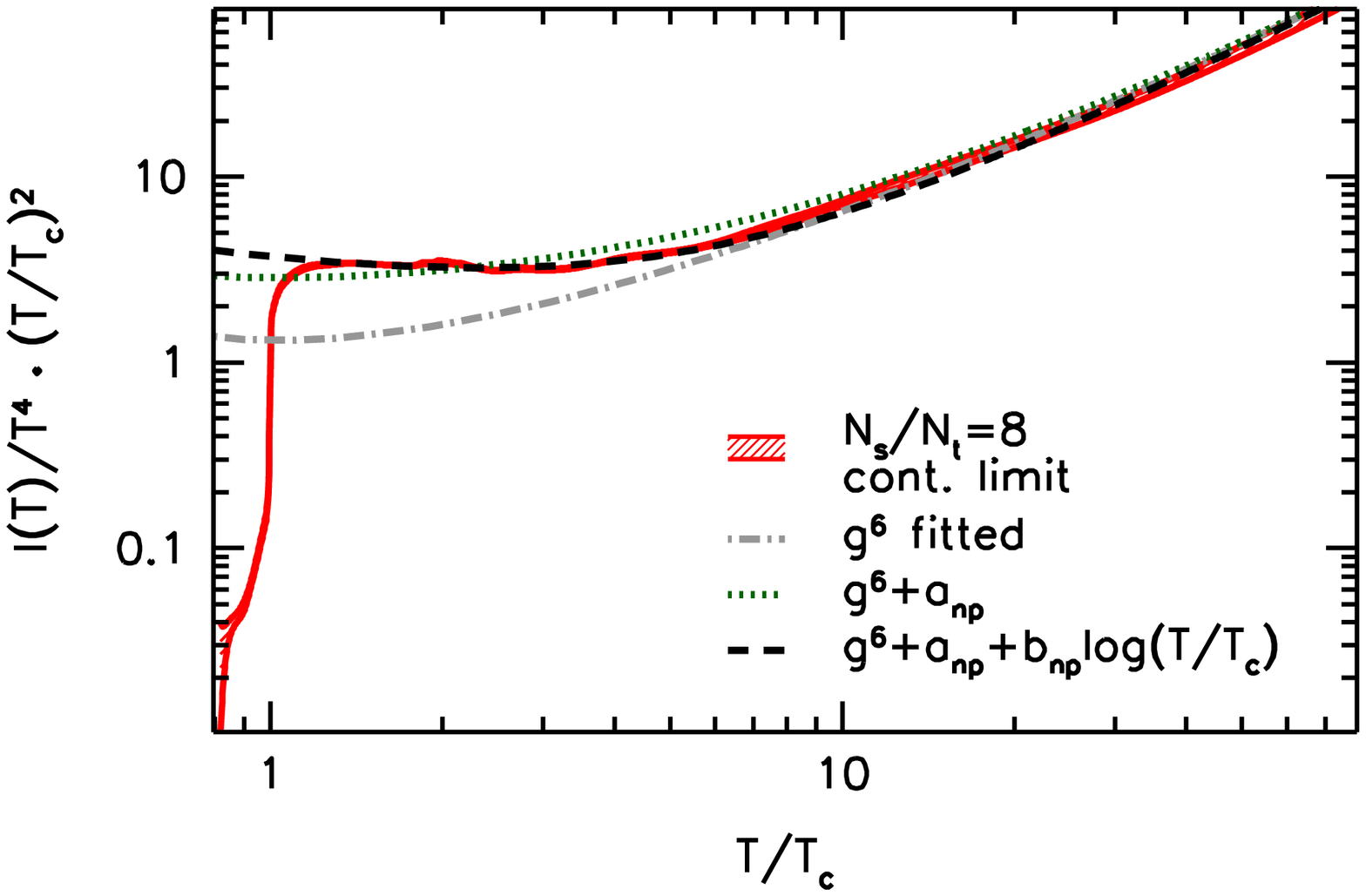,width=7.9cm,bb=18 360 592 718}
\epsfig{file=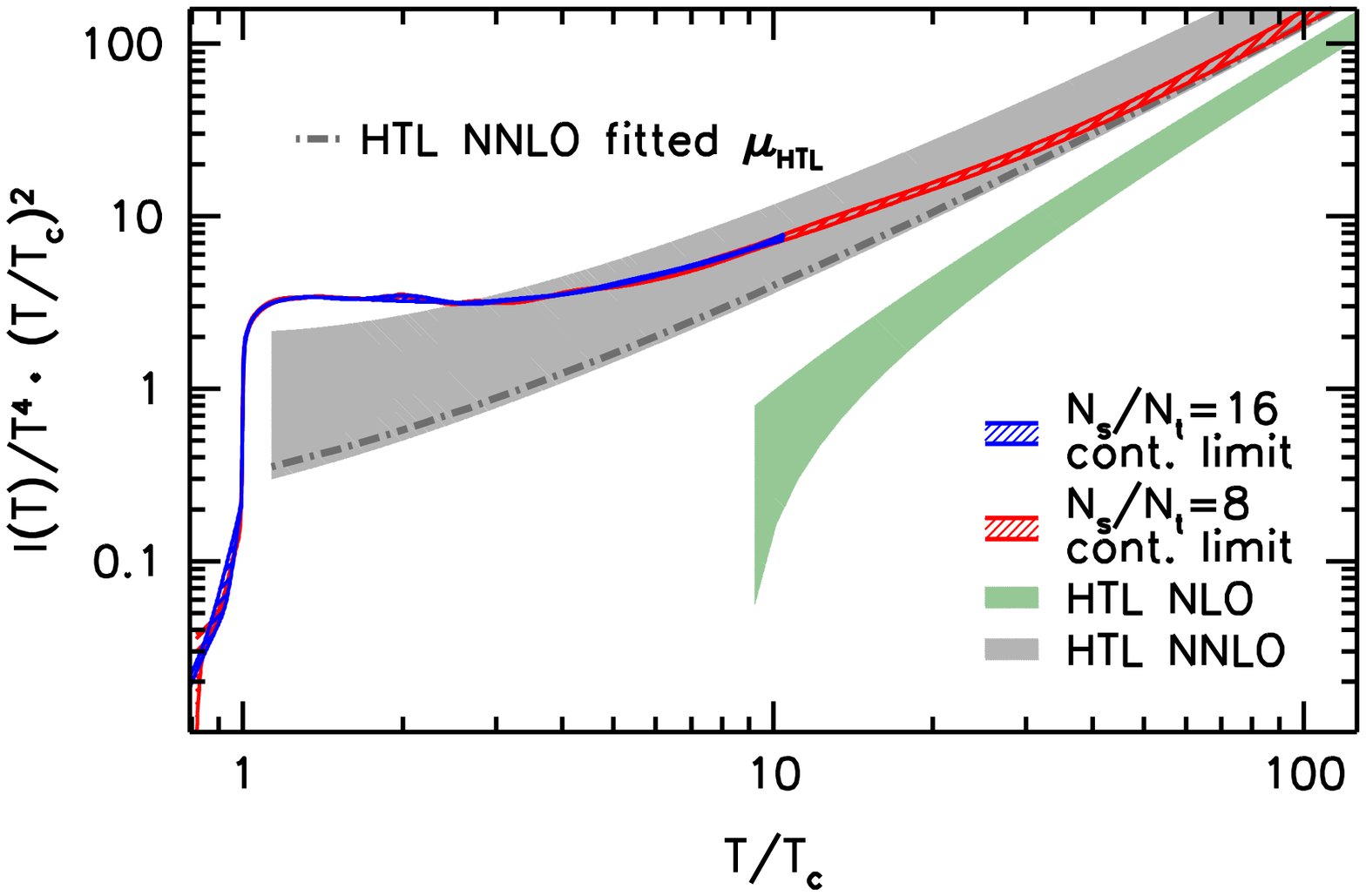,width=7.9cm,bb=18 360 592 718}
}
\caption{\label{fig:pertcomp}
{\bf Left side.} The continuum limit obtained from the lattice results (red band), compared to fitted perturbation theory. We fit the $g^6$ coefficient (gray dashed-dotted line), and the non-perturbative contribution in two different forms (black dashed line and green dotted line).
{\bf Right side.} The trace anomaly measured on two different spatial volumes (red and blue bands), compared to the NLO and NNLO HTL expansion with varied renormalization scale $0.5<\mu_{\rm HTL}/2\pi T<2$ (green and gray shaded regions). The dashed-dotted line represents the expansion with the fitted scale.
}
\vspace*{-0.3cm}
\end{figure}

Even in the small-volume simulations presented in this paper the lattices
do accommodate the ultrasoft scale $1/(g^2T)$. This encourages us to use these
very high temperature
lattice data to extract some information on the ultrasoft physics.  At high
enough temperatures, where the ${\cal O}(g^6)$ order is known to give a small
correction to the ${\cal O}(g^5)$ order one can fit the $g^6$ coefficient for
the perturbative result, and the preferred renomalization scale $\mu_{\rm HTL}$
for the HTL result. For the latter we find that our lattice data prefers
$\mu_{\rm HTL}/(2\pi T)=1.75(2)(6)(50)$, with the numbers in the parentheses
from left to right representing the statistical error, the error coming from
the lattice scale and that from the variation of the fit interval.  Repeating
the fit in Ref.~\cite{Kajantie:2002wa} we get for the sixth order coefficient
$q_c=-3526(4)(55)(30)$, in the same notation for the errors.  We used these
data in the left and right side of Fig\.~\ref{fig:pertcomp} to compare our
small volume simulations with the theoretical results.

\begin{figure*}
\centering
\hbox{
\epsfig{file=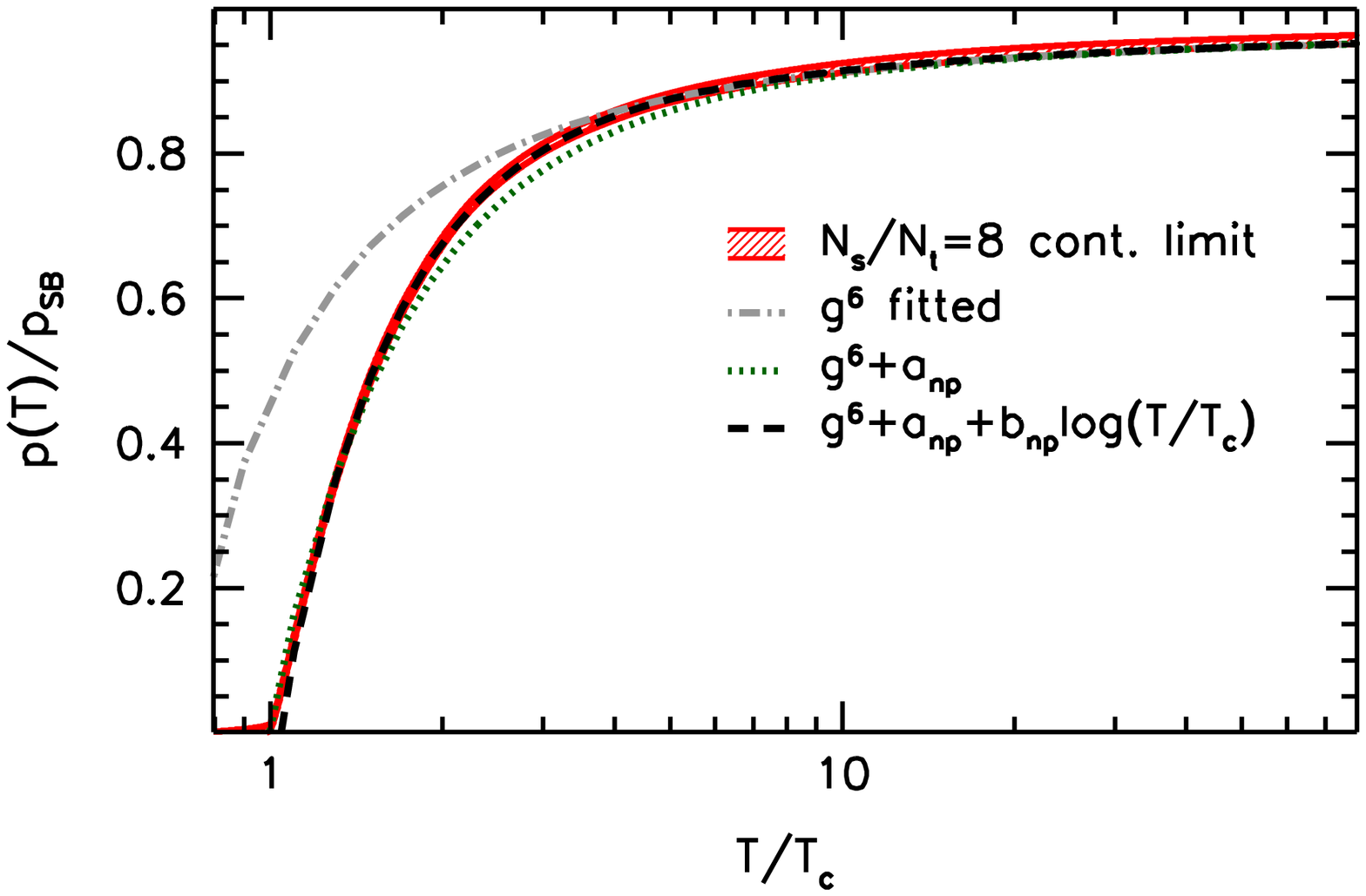,width=7.9cm,bb=18 360 592 718}
\epsfig{file=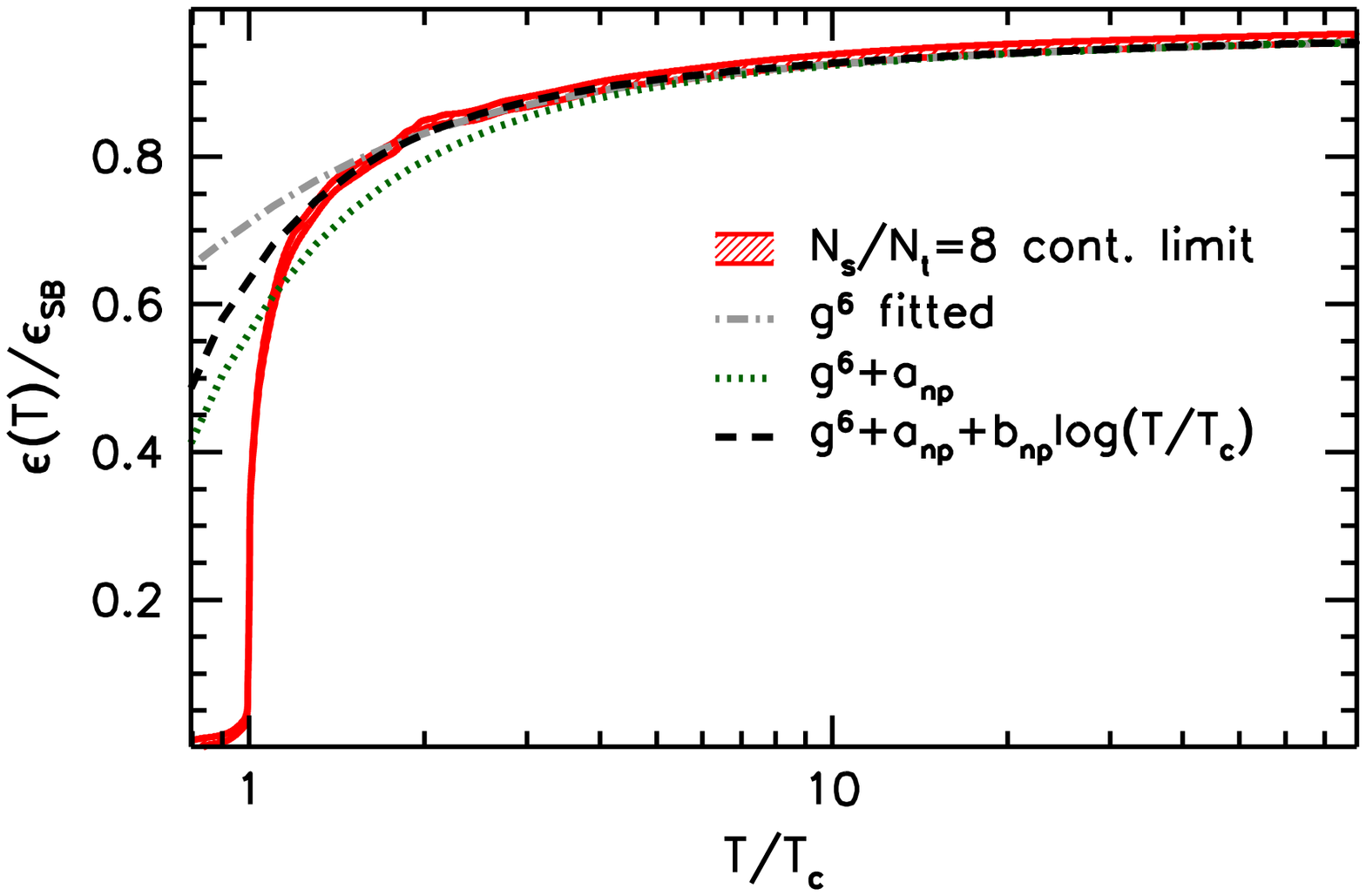,width=7.9cm,bb=18 360 592 718}
}
\caption{\label{fig:pertcompp}
The normalized pressure (left side) and energy density (right side) as measured on our small volume boxes. A comparison is shown to various fitted perturbative functions, in the same notation as in the left side of the previous figure.
}
\end{figure*}

In the figure we also plot the perturbative formula plus a nonperturbative $\sim T^{-2}$ contribution
using two simple functional forms $a_{\rm np}/T^2$ and $(a_{\rm np} + b_{\rm np} \log(T/T_c))/T^2$.
Using the former ansatz we obtain $a_{\rm np} = 0.879(2)(40)$, and the latter
$a_{\rm np} = 1.371(1)(50),\;b_{\rm np} = -0.618(2)(4)$. We note that these
parameters are rather sensitive to the variation of the lower endpoint of the
fit interval, but the relatively small $\chi^2/{\rm dof}\sim 4$ indicates (as
it can be seen in the figure) that the low-temperature region is well described
by the latter ansatz.  Using thermodynamic relations the fitted perturbative
formulae for the pressure and the energy density are also straightforward to
calculate. In figure~\ref{fig:pertcompp} we compare our small volume results to
the so obtained predictions. Similar comparisons can be made for the case of
the entropy density, too.

Tabulated data and the details of the present work will follow in a separate,
longer publication.

{\bf Acknowledgments:}
The simulations have mainly been performed on the QPACE facility.  The work was
partially supported by the DFG grants SFB-TR55 and FO-502/1-2.  Part of the
calculation was running on the GPU cluster at the E\"otv\"os University with
the support from the European Research Council grant 208740 (FP7/2007-2013).
The authors acknowledge the helpful comments from Axel Maas, Aleksi Kurkela,
Marco Panero, Kari Rummukainen and York Schr\"oder as well as the fruitful
correspondence with Mike Strickland.

\end{document}